\documentclass[11pt]{article} %for pdfLaTeX
\usepackage{cite}
\usepackage{latexsym}
\usepackage{amssymb}
\usepackage{amsmath}
\usepackage[breaklinks=true,linktocpage=true]{hyperref}
\hypersetup{colorlinks=true, allcolors=blue}
\usepackage{graphicx}% Include figure files

\usepackage{slashed}
\usepackage{mathbbol}

\usepackage{color}
\usepackage{comment}
\usepackage{physics}
\usepackage{float}
\usepackage{siunitx}

\begin{document}
\pagestyle{empty}

\begin{flushright}
KEK-TH-2638
\end{flushright}

\vspace{3cm}

\begin{center}

{\bf\LARGE
T violation at a future neutrino factory
}
\\

\vspace*{1.5cm}
{\large 
Ryuichiro Kitano$^{1,2}$, 
Joe Sato$^{3}$, and 
Sho Sugama$^{3}$
} \\
\vspace*{0.5cm}

{\it 
$^1$KEK Theory Center, Tsukuba 305-0801,
Japan\\
$^2$Graduate University for Advanced Studies (Sokendai), Tsukuba
305-0801, Japan\\
$^3$Department of Physics, Facility of Engineering Science, Yokohama National University, Yokohama 240-8501, Japan 
}

\end{center}

\vspace*{1.0cm}

\begin{abstract}
{\normalsize
We study the possibility of measuring T (time reversal) violation
in a future long
baseline neutrino oscillation experiment. By assuming a neutrino
factory as a staging scenario of a muon collider at the J-PARC site,
we find that the $\nu_e \to \nu_\mu$ oscillation probabilities can be
measured with a good accuracy at the Hyper-Kamiokande detector. 
By comparing with the probability of the time-reversal process,
$\nu_\mu \to \nu_e$, measured at the T2K/T2HK experiments, one can
determine the CP phase $\delta$ in the neutrino mixing matrix if
$|\sin(\delta)|$ is large enough. 
The determination of $\delta$ can be made with poor knowledge of the
matter density of the earth as T violation is almost insensitive to the
matter effects. The comparison of CP and T-violation measurements,
{\it \`a la} the CPT theorem, provides us with a non-trivial check of
the three neutrino paradigm based on the quantum field theory.
}
\end{abstract} 

%%%%%%%%%%%%%%%%%%%%%%%%%%%%%%%%%%%%%%%%%%%%%%%%%%%%%%%%%%%%%%%%%%%%%%%%%%%%
\newpage
\baselineskip=18pt
\setcounter{page}{2}
\pagestyle{plain}
\baselineskip=18pt
\pagestyle{plain}

\setcounter{footnote}{0}

\section{Introduction}

The neutrino oscillation phenomena have provided us with the largest
scale tests of the time evolution of physical states via quantum
mechanics. The terrestrial experiments such as T2K~\cite{T2K:2021xwb} and NO$\nu$A~\cite{NOvA:2021nfi}
experiments are now operating to look for CP violating phenomena by
using $O(100-1000)$~km size baselines.
In future, such as within the time scale of ten years,
Hyper-Kamiokande~\cite{Hyper-Kamiokande:2018ofw}, JUNO~\cite{JUNO:2015zny} and DUNE~\cite{DUNE:2015lol} experiments are expected to uncover
the oscillation parameters in the three neutrino scheme, such as the
ordering of the neutrino masses and the size of the Dirac CP phase.

Once we know all of those, we can further ask interesting questions.
Do neutrinos obey usual rules in the quantum mechanics? Do neutrinos
have only weak interactions? Are there only three neutrino
generations?
It will be important to over-constrain the oscillation parameters to
check if we really understand the lepton sector of the Standard Model,
just as we have done to confirm the Kobayashi-Maskawa theory in the
quark sector.

Of course, in the current situation, the discovery of CP violation in
the neutrino sector itself will be the most important milestone. The
standard method is to look for the difference in the $\nu_\mu \to
\nu_e$ and $\bar \nu_\mu \to \bar \nu_e$ oscillations in the
accelerator based long-baseline experiments~\cite{K2K:2006yov}. Although observing the
difference sounds a clear CP-violating signal, matter effects in the
neutrino propagation in the earth make the analysis complicated as
we need an anti-earth for the difference to be formally CP-odd. In
other words, imperfect knowledge of the matter-density profile of the
earth can fake the CP violation.

The importance of measuring T violation has been pointed out in
Ref.~\cite{Cabibbo:1977nk,Kuo:1987km,Krastev:1988yu,Toshev:1989vz,Toshev:1991ku,Arafune:1996bt,Arafune:1997hd,Koike:1999tb}.
Many physicists, say, in Ref~\cite{Koike:1999hf,Koike:2000jf,Miura:2001pia,Pinney:2001xw,Akhmedov:2001kd,Miura:2001pi,Ota:2001cz,Bueno:2001jd} considered 
the use of high intensity muon beams, so called 
neutrino factory~\cite{Geer:1997iz}. The collimated $\nu_e$ flux from the
muon decays in flight can be a good source for long baseline
oscillation experiments.
The observation of the $\nu_e \to \nu_\mu$ oscillation will be made
possible in such experiments, and by comparing with $\nu_\mu \to
\nu_e$, one can measure T violation.
The virtue of T violation is that it is insensitive to the matter
effects. Formally, the T violation is the difference between the
oscillation probabilities of $\nu_\mu \to \nu_e$ and $\nu_e \to
\nu_\mu$ while flipping the locations of source and the detector. By
approximating the earth as a spherically symmetric object, the
flipping would not change the probabilities, and thus one can perform
the experiments by the same facility of the $\nu_\mu \to \nu_e$
measurements such as T2K and T2HK.
It is also interesting to compare CP-violation measurements with
T-violation ones. This is going to be a very non-trivial consistency
check of the quantum mechanics of the three-neutrino paradigm.

Recently, neutrino-factory programs have been revived as a staging
scenario of muon colliders~(see for e.g.,~\cite{Kaplan:2012zzb}).
In Japan, there is a proposal to utilize the ultra slow muon
technology~\cite{Kondo:2018} for muon cooling that is the essential part to realize a
muon collider~\cite{Hamada:2022mua}.
For example, one can use J-PARC as the proton driver to produce pions,
and by catching almost all the pions as well as the muons from the
pion decays, one can obtain $O(10^{14})$ cold positive muons per second.
With this number, one can consider a high energy and high intensity
$\mu^+ e^-$ and $\mu^+\mu^+$ colliders, which are excellent Higgs
boson factories.
As one of the staging scenarios, one can consider a low energy $\mu^+$
beam such as $O(1)$~GeV for neutrino factory to shoot $\nu_e$ beams to
Super/Hyper-Kamiokande.
Indeed, the ultra slow muon technology has been used in the 
muon $g-2$/EDM experiment at J-PARC~\cite{Abe:2019thb} to obtain 
a low-emittance and polarized muon beam which is to be 
accelerated up to about 200~MeV within the time scale 
of a few years.

In this paper, we study the possibility of measuring T violation in
the long baseline experiments. As an example, we take the same
baseline as the T2HK experiments, and assume the neutrino flux
obtained from the number of muons mentioned above.
We conduct $\chi^2$ analyses to evaluate the possibility of testing T violation. 

This paper is organized as follows. In Section~\ref{sec:osc}, we
review neutrino oscillations in the three-flavor scheme, and discuss T
and CP violation in neutrino oscillations. In Section~\ref{sec:chi2},
we conduct a statistical analysis to estimate the sensitivity of the
long baseline experiment based on the $\nu_e$ beam from the neutrino
factory to T violation defined as the probability difference.
Section~\ref{sec:summary} is devoted to summary.

%\newpage
\section{CP and T violation in neutrino oscillations}
\label{sec:osc}

We explain here our notations used in this paper and derive formulae
for CP and T violation in the oscillation probabilities. We present
the energy dependencies of them and discuss the impacts of the matter
effects.

We parametrize below the PMNS matrix which is a $3 \times 3$ unitary
matrix to relate the mass and flavor eigenstates in the lepton sector:
\begin{align}
    U &= \mqty(U_{e1}&U_{e2}&U_{e3} \\ U_{\mu1}&U_{\mu2}&U_{\mu3} \\ U_{\tau1}&U_{\tau2}&U_{\tau3}) \label{eq1} \\
    &=\mqty(c_{12}c_{13}&s_{12}c_{13}&s_{13}e^{-i\delta} \\
             -s_{12}c_{23}-c_{12}s_{23}s_{13}e^{i\delta}&c_{12}c_{23}-s_{12}s_{23}s_{13}e^{i\delta}&s_{23}c_{13} \\
              s_{12}s_{23}-c_{12}c_{23}s_{13}e^{i\delta}&-c_{12}s_{23}-s_{12}c_{23}s_{13}e^{i\delta}&c_{23}c_{13}),
              \label{eq2}
\end{align}
with the notation, $s_{ij} = \sin \theta_{ij}$ and $c_{ij} = \cos \theta_{ij}$.
The time evolution of neutrinos is described by the following
Schr\"odinger equation,
\begin{align}
    i\dv{}{t}\mqty(\nu_e(\bar{\nu}_e)\\ \nu_\mu(\bar{\nu}_\mu) \\ \nu_\tau(\bar{\nu}_\tau)) 
    %&=\left[U^{(*)} \mathrm{diag}(0,\mathit{\Delta} E_{21},\mathit{\Delta} E_{31})U^{\dag(T)}+\mathrm{diag}(\pm A,0,0)\right]\mqty(\nu_e(\bar{\nu}_e)\\ \nu_\mu(\bar{\nu}_\mu) \\ \nu_\tau(\bar{\nu}_\tau))\notag \\
    &= \mathcal{M}^{(\pm)}\mqty(\nu_e(\bar{\nu}_e)\\ \nu_\mu(\bar{\nu}_\mu) \\ \nu_\tau(\bar{\nu}_\tau)),
    \label{eq3}
\end{align}
where we combined the equations for neutrinos ($\nu$) and
anti-neutrinos ($\bar \nu$).
The effective Hamiltonian in matter is given by
\begin{align}
    \mathcal{M}^{(\pm)} \equiv U^{(*)}\mathrm{diag}(0,\mathit{\Delta} 
    E_{21},\mathit{\Delta} E_{31})U^{\dag(T)}+\mathrm{diag}(\pm A,0,0).
    \label{eq4}
\end{align}
where $\mathit{\Delta}E_{jk} \equiv \mathit{\Delta}m_{jk}^2/2E$ are
the differences in the energy eigenvalues in vacua and $A =
\sqrt{2}G_F n_e$ represents the matter effect. Here, the 
density of the electron in the earth, $n_e$, can be calculated 
from the matter density of the earth, $\rho$, which 
is in our analysis taken to be constant along the baseline for simplicity~\cite{Dziewonski:1981xy}. 
From this equation, we obtain expressions of the neutrino oscillation
probabilities for a baseline length $L$ as
\begin{align}
    P(\nu_\alpha(\bar{\nu}_\alpha) \to \nu_\beta(\bar{\nu}_\beta)) = \delta_{\alpha\beta}
    & -4\sum_{j>k}\mathrm{Re}\left[\tilde{U}^{(\pm)}_{\alpha j}\tilde{U}^{(\pm)*}_{\beta j} \tilde{U}^{(\pm)*}_{\alpha k} \tilde{U}^{(\pm)}_{\beta k}\right]\sin^2\left(\frac{\mathit{\Delta} \tilde{E}^{(\pm)}_{jk}L}{2}\right)\notag\\
    & +2\sum_{j>k}\mathrm{Im}\left[\tilde{U}^{(\pm)}_{\alpha j}\tilde{U}^{(\pm)*}_{\beta j} \tilde{U}^{(\pm)*}_{\alpha k} \tilde{U}^{(\pm)}_{\beta k}\right]\sin\left(\mathit{\Delta} \tilde{E}^{(\pm)}_{jk}L\right)
    \label{eq5}
\end{align}
where $\tilde{U}^{(\pm)}$ is the matrix diagonalizing
$\mathcal{M}^{(\pm)}$ and
$\tilde{E}^{(\pm)}_{jk}=\tilde{E}^{(\pm)}_j-\Tilde{E}^{(\pm)}_k$,
$\Tilde{E}^{(\pm)}_j$ are eigenvalues of $\mathcal{M}^{(\pm)}$.

The CP violation in the neutrino oscillation can be defined as
\begin{align}
    & P\qty(\nu_\alpha \to \nu_\beta) - P\qty(\bar{\nu}_\alpha \to \bar{\nu}_\beta) \notag \\
    &= -4\sum_{j>k}\mathrm{Re} \left[ \tilde{U}^{(+)}_{\alpha j}\tilde{U}^{(+)*}_{\beta j} \tilde{U}^{(+)*}_{\alpha k} \tilde{U}^{(+)}_{\beta k}\sin^2\left(\frac{\mathit{\Delta} \tilde{E}^{(+)}_{jk}L}{2}\right) - \tilde{U}^{(-)}_{\alpha j}\tilde{U}^{(-)*}_{\beta j} \tilde{U}^{(-)*}_{\alpha k} \tilde{U}^{(-)}_{\beta k}\sin^2\left(\frac{\mathit{\Delta} \tilde{E}^{(-)}_{jk}L}{2}\right) \right] \notag \\
    & +2\sum_{j>k}\mathrm{Im}\left[\tilde{U}^{(+)}_{\alpha j}\tilde{U}^{(+)*}_{\beta j} \tilde{U}^{(+)*}_{\alpha k} \tilde{U}^{(+)}_{\beta k}\sin\left(\mathit{\Delta} \tilde{E}^{(+)}_{jk}L\right) - \tilde{U}^{(-)}_{\alpha j}\tilde{U}^{(-)*}_{\beta j} \tilde{U}^{(-)*}_{\alpha k} \tilde{U}^{(-)}_{\beta k}\sin\left(\mathit{\Delta} \tilde{E}^{(-)}_{jk}L\right)\right]
    \label{eq6}
\end{align}
It is important to note that, even if the CP phase,
$\delta$, is zero, this quantity does not necessarily vanish. This is due to the
fact that CP transformation of matter should be anti-matter, and thus
the quantity defined above is not quite CP-odd.

On the other hand, T violation can be defined as
\begin{align}
    P\qty(\nu_\alpha \to \nu_\beta) - P\qty(\nu_\beta \to \nu_\alpha) \notag 
    &= 4\sum_{j>k}\mathrm{Im}\left[ \tilde{U}^{(+)}_{\alpha j}\tilde{U}^{(+)*}_{\beta j} \tilde{U}^{(+)*}_{\alpha k} \tilde{U}^{(+)}_{\beta k} \right]\sin\left(\mathit{\Delta} \tilde{E}^{(+)}_{jk}L\right) \notag \\
    &=4\tilde{J}^{(+)} \left[\sin\qty(\mathit{\Delta} \tilde{E}^{(+)}_{12}L) + \sin\qty(\mathit{\Delta} \tilde{E}^{(+)}_{23}L) + \sin\qty(\mathit{\Delta} \tilde{E}^{(+)}_{31}L) \right] \notag \\
    &= -16 \tilde{J}^{(+)} \sin\qty(\frac{\mathit{\Delta}\tilde{E}^{(+)}_{31}L)}{2}) \sin\qty(\frac{\mathit{\Delta}\tilde{E}^{(+)}_{32}L)}{2}) \sin\qty(\frac{\mathit{\Delta}\tilde{E}^{(+)}_{21}L)}{2})
    \label{eq7}
\end{align}
where $\tilde{J}^{(+)} \equiv \mathrm{Im}\left[
\tilde{U}^{(+)}_{\alpha 1}\tilde{U}^{(+)*}_{\beta 1}
\tilde{U}^{(+)*}_{\alpha 2} \tilde{U}^{(+)}_{\beta 2} \right]$ is the
modified Jarlskog factor in
matter~\cite{Harrison:1999df,Naumov:1991rh,Naumov:1991ju}.
It is known that this factor can be expressed in terms of the Jarlskog invariant in the vacuum, $J = s_{13} c_{13}^2 s_{12} c_{12} s_{23} c_{23} \sin \delta$~\cite{Jarlskog:1985cw,Jarlskog:1985ht},
as 
\begin{align}
    \tilde{J}^{(\pm)} = \frac{\mathit{\Delta}E_{31}\mathit{\Delta}E_{32}\mathit{\Delta}E_{21}}{\mathit{\Delta}\tilde{E}^{(\pm)}_{31}\mathit{\Delta}\tilde{E}^{(\pm)}_{32}\mathit{\Delta}\tilde{E}^{(\pm)}_{21}}J.
    \label{eq8}
\end{align}
T violation here is formally T-odd when we assume that the matter
density is symmetric under the exchange of the neutrino source and the
detector.
Consequently, if $\delta$ is zero, the difference $P\qty(\nu_\alpha
\to \nu_\beta) - P\qty(\nu_\beta \to \nu_\alpha)$ is also exactly
zero. Therefore, when measuring $\delta$, T violation gives clearer
signals~\cite{Krastev:1988yu,Arafune:1996bt,Akhmedov:2001kd}. 

\begin{figure}[h]
    \centering
    \includegraphics[width=15cm]{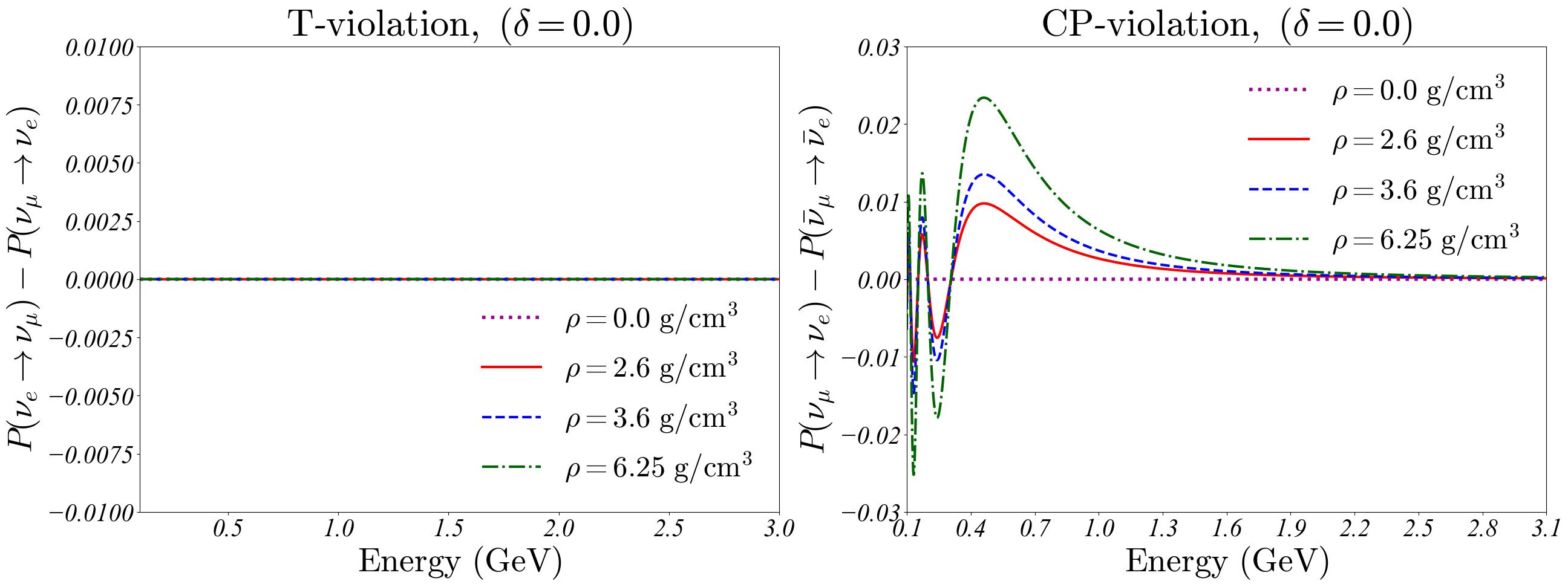}
    \caption{Energy dependence of T and CP violation for various choice
    of matter densities. The CP phase in the PMNS matrix is taken to be $\delta = 0$. The purple dotted, red solid, blue dashed, and green dashdot lines correspond to $0.0~\rm{g/cm^3}$, $2.6~\rm{g/cm^3}$, $3.6~\rm{g/cm^3}$, $6.25~\rm{g/cm^3}$, respectively.}
    \label{fig2}
\end{figure}
\begin{figure}[h]
    \centering
    \includegraphics[width=15cm]{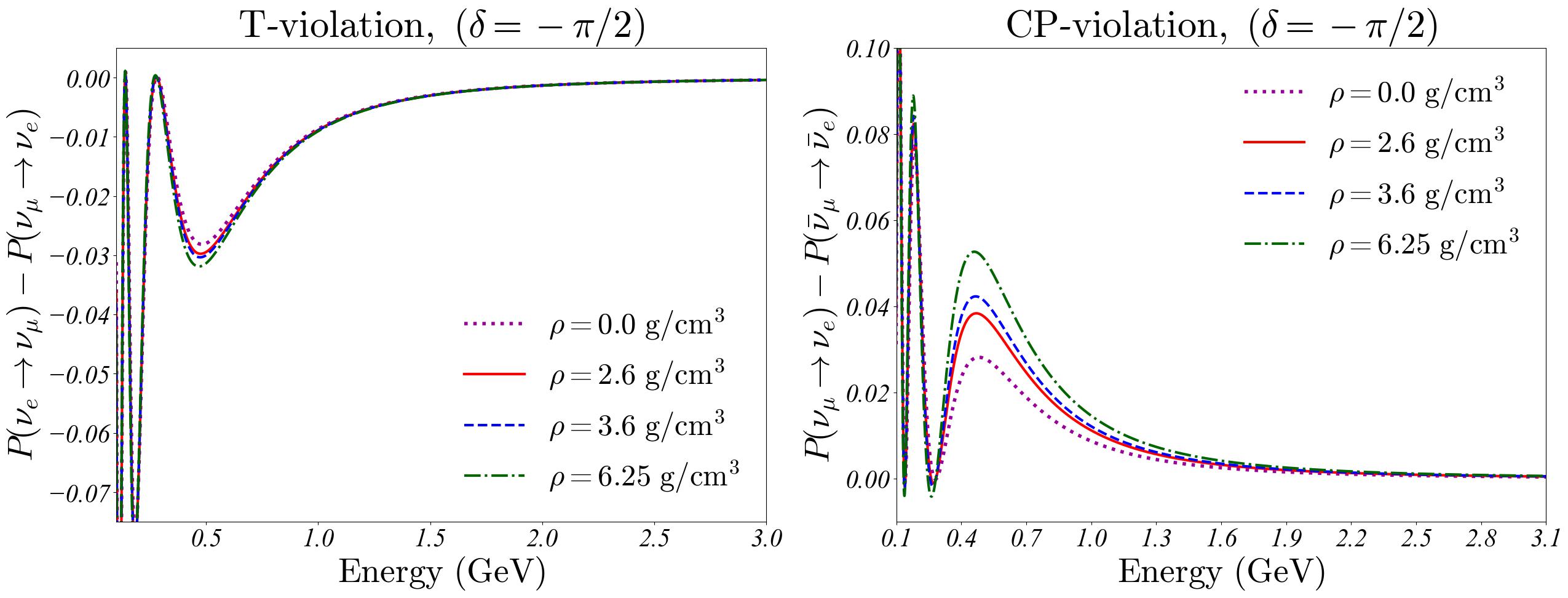}
    \caption{Energy dependence of T and CP violation for various choice
    of matter densities. The CP phase in the PMNS matrix is taken to be $\delta = -\pi/2$.}
    \label{fig1}
\end{figure}

As a demonstration, we solve Eq.~\eqref{eq3} for different matter
densities, $\rho$, and show in Fig.~\ref{fig2} the energy dependence of the T
and CP violation defined above for the case of $\delta = 0$. It is
clear that T violation vanishes while we see non-vanishing
``CP violation,'' which depend on the matter density $\rho$.
This means, in order to establish the true CP violation, i.e., $\delta
\neq 0$ nor $\pi$, via the measurements of the ``CP violation,'' we need good
knowledge of matter profile in the earth.
The same figure for $\delta = -\pi/2$ is shown in Fig.~\ref{fig1}. We
see a large dependence on the matter density in the case of
CP violation, while T violation has essentially no difference.
Although a little dependence shows up through the $\rho$ dependence
in Eq.~\eqref{eq8}, one can see that complicated matter effects are
largely removed in T violation measurements.

%
%
%%%%%%%%%%%%%%%%%%%%%%%%%%%%%%%%%%%%%%%%%%%%%%%%%%%%%%%%%%%%%%%%%%%%%
%\newpage
\section{Statistical analysis}
\label{sec:chi2}

We discuss the statistical precisions of the CP and T violation
measurement defined in the previous section. We assume a race-track type 
storage ring of the 
high-intensity muon beam at the J-PARC site with one of the straight
section pointing towards the Hyper-Kamiokande detector.
The decay of the beam muon produces a highly collimated neutrino beam
and thus such a facility is referred-to as a neutrino factory~\cite{Geer:1997iz}.
Motivated by the discussion of muon collider experiments using the
ultra slow muons ($\mu$TRISTAN), we assume a $\mu^+$ beam with a
monochromatic energy at 1.5~GeV. The produced neutrinos are,
therefore, $\bar \nu_\mu$ and $\nu_e$, with some energy distributions,
which depends on the polarization of $\mu^+$.
According to the estimates given in Ref.~\cite{Hamada:2022mua}, the number
of generated cold muons can be as large as $10^{14}$ per second
even after considering all the efficiencies for muon stopping and laser ionizations.
We expect that one can generate a few times $10^{21}$ per year in this 
type of facility.
Since we assume a race-track type muon storage ring with two straight sections, we can 
make use of roughly 1/3 of muons whose decay products point to the far detector~\cite{Barger:1999fs}.
We expect that we are able 
to use $N_\mu = 10^{21}/$year, and the total running time to be of the order of an year.

For the T-violation measurement, we use the $\nu_e$ beam to measure
the oscillation probability $P(\nu_e \to \nu_\mu)$ by comparing the
number of $\nu_e$ produced at J-PARC and that of $\nu_\mu$ detected at
Hyper-Kamiokande. The time-reversal probability, $P(\nu_\mu \to
\nu_e)$ is measured at the T2HK experiment by using the conventional
beam from the pion ($\pi^+$) decays. We assume the same baseline for
these two experiments. The CP violation, $P(\nu_\mu \to \nu_e) -
P(\bar \nu_\mu \to \bar \nu_e)$ is also measured at the T2HK
experiment.

For the measurement of $P(\nu_e \to \nu_\mu)$, we need to distinguish
$\nu_\mu$ from $\bar \nu_\mu$ at Hyper-Kamiokande as $\bar \nu_\mu$ is
also produced by the decay of $\mu^+$.
The water Cherenkov detector such as Hyper-Kamiokande can, in
principle, identify the charge of the muons generated by the charged
current process~\cite{Beacom:2003nk,Huber:2008yx} , $\nu_\mu n \to \mu^- p$ and $\bar \nu_\mu p \to \mu^+ 
n$ by tagging the neutron in the final state, but the efficiencies of 
such signals are currently under
studies~\cite{Beacom:2003nk,Super-Kamiokande:2023xup,Akutsu:2019} .
We will discuss how we extract the oscillation probabilities under the
$\bar \nu_\mu$ background later in this section.

\subsection{Number of events at Hyper-Kamiokande}
\begin{figure}[H]
    \centering
    \includegraphics[width=15cm]{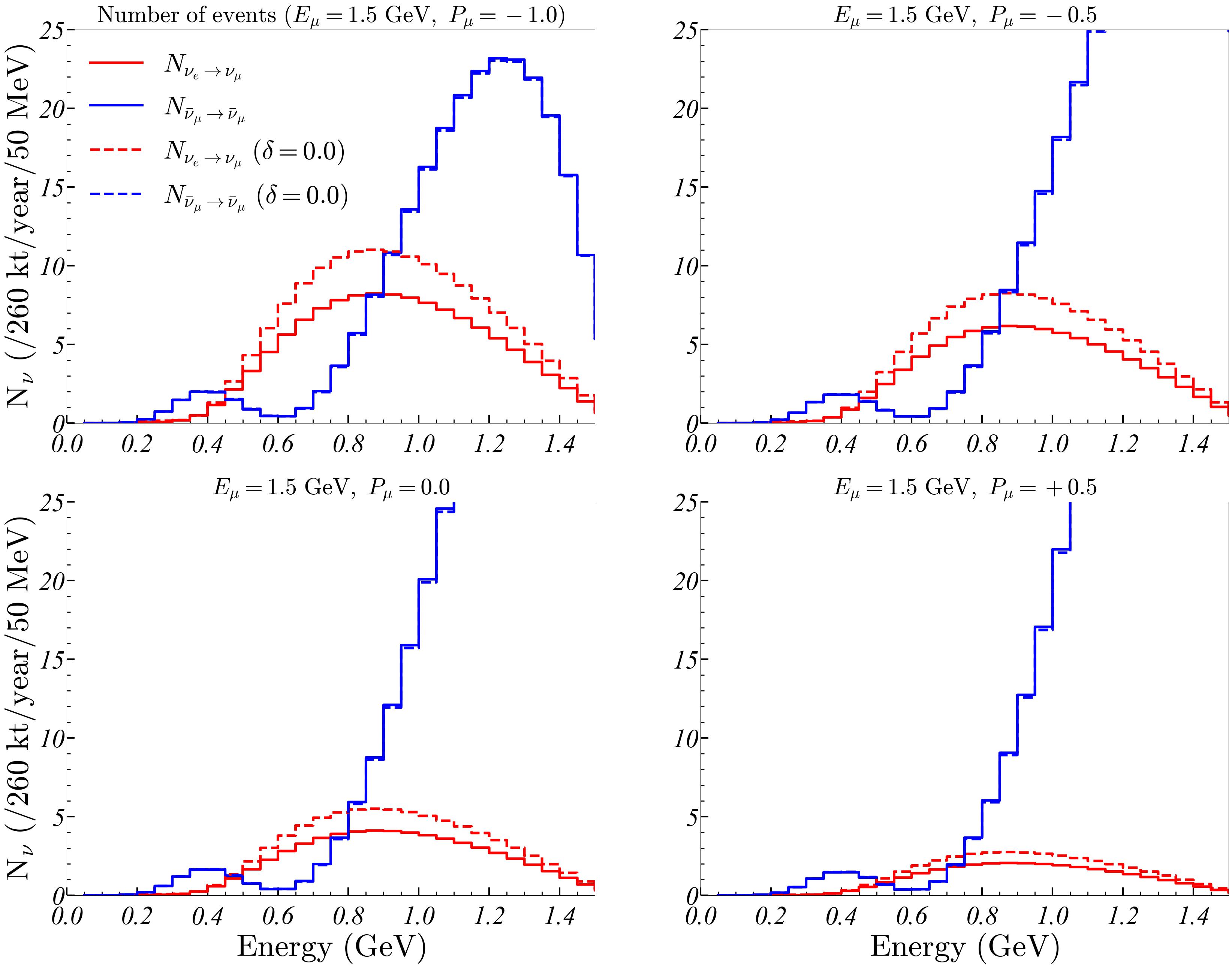}
    \caption{Neutrino flux measured at the far detector. The anti-muon beam energy is set to $1.5~\mathrm{GeV}$, and calculations are performed for four polarizations ($P_\mu=-1.0,\ -0.5,\ 0.0,\ +0.5$). The solid (dashed) lines depict the fluxes for $\delta=-\pi/2$ ($\delta=0.0$).}
    \label{fig3}
\end{figure}

The neutrino flux from the neutrino factory can be estimated by the
neutrino-energy ($E_\nu$) distribution of the (polarized) $\mu^+$
decay with a fixed energy $E_\mu$~\cite{Barger:1999fs}. By convoluting
the $E_\nu$ distribution with the oscillation probabilities, $P^{\nu_e
\to \nu_\mu} (E_\nu)$, and the cross section of the neutrino-nucleon
scattering, $\sigma_{\nu_\mu} (E_\nu)$, one can obtain the event rate
in each energy bin. The number of events is given by
\begin{align}
    N_j^{\nu_e \to \nu_\mu} &= \int_{E_j}^{E_{j+1}} \frac{dE_\nu}{E_\mu} \times \frac{12N_\mu\cdot V\cdot n_N}{\pi L^2} \times \gamma^2 \qty(\frac{E_\nu}{E_\mu})^2 \times \qty[\qty(1-\frac{E_\nu}{E_\mu}) -P_\mu\qty(1-\frac{E_\nu}{E_\mu})] \notag \\
    & \times P^{\nu_e\to\nu_\mu}(E_\nu) \times \sigma_{\nu_\mu}(E_\nu).
    %
    %  \notag \\
    % & = \frac{12N_\mu\cdot V_{kt} \cdot N_A}{\pi L^2} \times \frac{1}{m_\mu^2E_\mu} \times \int_{E_j}^{E_{j+1}}dE_\nu \times E_\nu^2\qty(1-P_\mu\cos\theta)\qty(1-\frac{E_\nu}{E_\mu}) \notag \\
    % &\times P^{\nu_e\to\nu_\mu}(E_\nu) \times \sigma_{\nu_\mu}(E_\nu)
    \label{eq9}
\end{align}
where $V$ is the detector volume, $n_N$ is the number density of the nucleon in water,
and $L$ is the length of the baseline. The boost factor $\gamma$ is
that for the muon, i.e., $\gamma = E_\mu / m_\mu$ with $m_\mu$ the
muon mass.
We assume a polarization of the anti-muon beam $P_\mu$ which is possible in
the scheme of the ultra slow muon. The polarization will be important
to obtain a larger $\nu_e$ flux in the forward region.
Since we are going to perform combined analyses with the T2HK experiment,
the far detector is assumed to be Hyper-Kamiokande and the baseline
length is $L = 295\ \mathrm{km}$. 
We show in Fig.~\ref{fig3} the number of events, $N_j^{\nu_e \to
\nu_\mu}$ for a choice of the CP-phase $\delta = -\pi/2$ (red solid)
and $\delta = 0$ (red dashed) for $P_\mu = -1.0$, $-0.5$,
$0.0$, and $+0.5$ for anti-muon. 
Here we smeared the neutrino spectrum by a representative value of the detector resolution, $\Delta E = 50$~MeV~Ref.\cite{Hayato:2021heg}. We have confirmed that the choices of $\Delta E = 0$~MeV or $85$~MeV would hardly ever modify the results we show in this paper, as we anyway analyze binned data by 50~MeV. 
We also overlaid the number of background events, $N_j^{\bar \nu_\mu \to
\bar \nu_\mu}$ (blue solid and dashed).
Near the energy region $E_\nu \sim 0.5 - 0.7$~GeV, the background events are suppressed
since that is the energy for the oscillation maximum for $L=295$~km.
This is quite fortunate situation as the signal events (red) have a
peak near such energies.
We also see that the number of events are the largest for $P_\mu = -1.0$.
At this stage, one can expect that the measurement of the $\nu_e \to
\nu_\mu$ process will be possible with a neutrino factory at J-PARC
with $N_\mu = 10^{21}$/year.
We use the cross section data for charged current quasielastic interactions from Ref.~\cite{Formaggio:2012cpf}.

%fig of neutrino flux

\begin{figure}[h]
    \centering
    \begin{tabular}{cc}
        \begin{minipage}[t]{0.49\hsize}
            \centering
            \includegraphics[width=7.5cm]{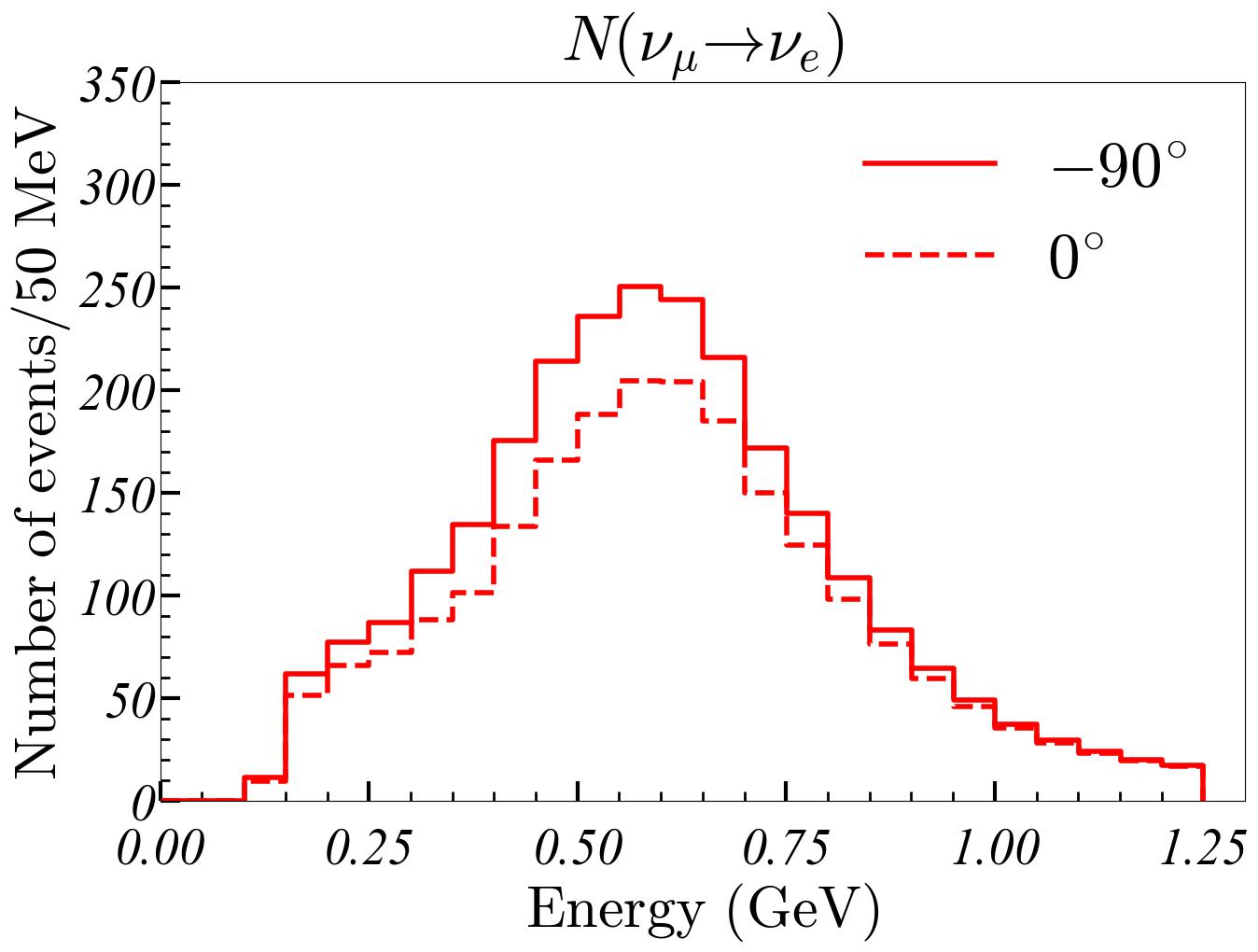}
        \end{minipage} &
        \begin{minipage}[t]{0.49\hsize}
            \centering
            \includegraphics[width=7.5cm]{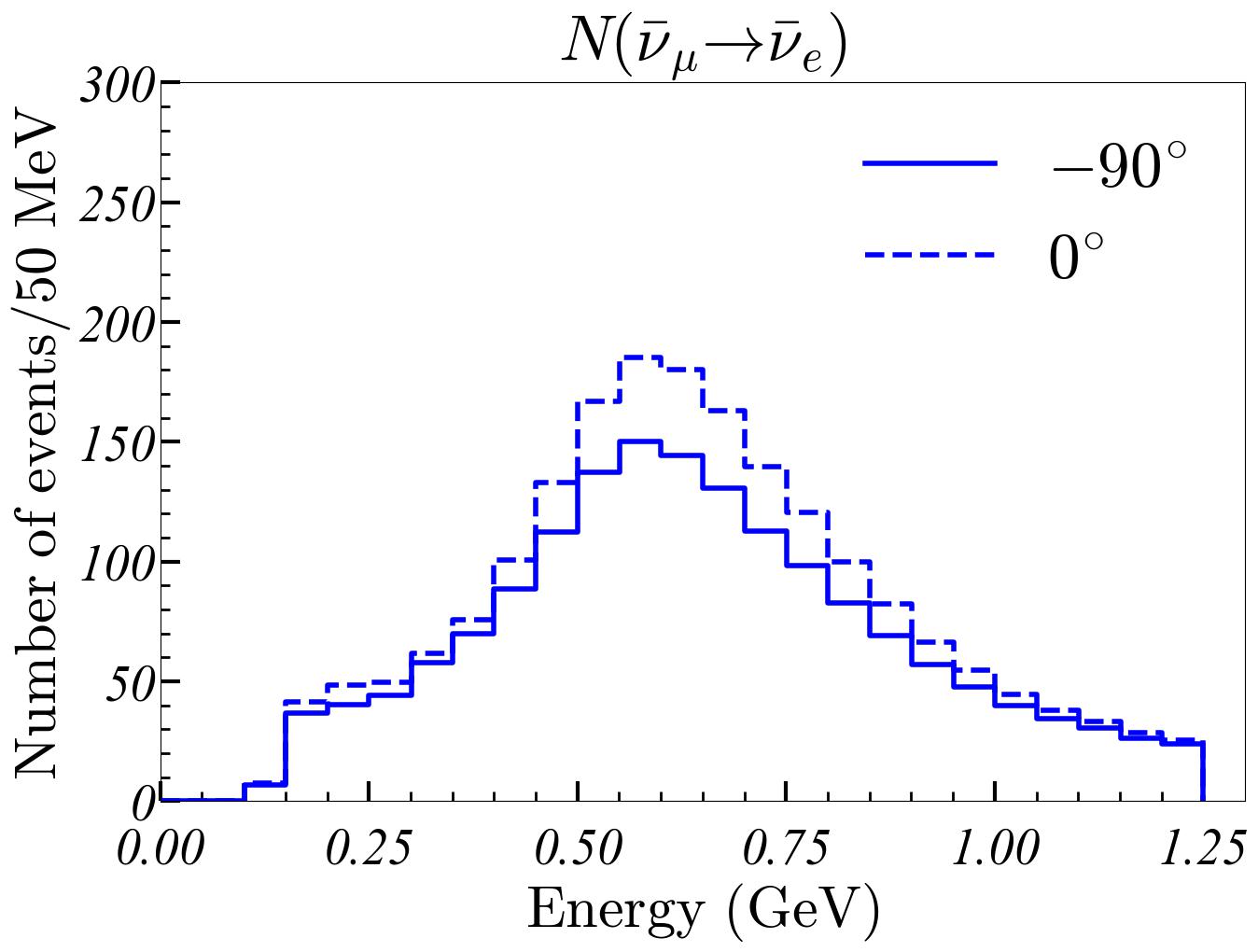}
        \end{minipage} 
    \end{tabular}
    \caption{Number of events in the T2HK experiment. The left and right figures show the fluxes of $\nu_\mu \to \nu_e$ and $\bar{\nu}_\mu \to \bar{\nu}_e$ at the far detector, respectively. The solid (dash) lines correspond to $\delta=\ang{-90}\ (\delta=\ang{0})$. These expected number of events are taken from Ref.~\cite{Hyper-Kamiokande:2018ofw}.}
    \label{fig4}
\end{figure}

We also show in Fig.~\ref{fig4} the expected number of events for the
CP-violation measurements, $N_j^{\nu_\mu \to \nu_e}$ and $N_j^{\bar
\nu_\mu \to \bar \nu_e}$ in the T2HK experiment. We use the neutrino
flux calculated in Ref.~\cite{Hyper-Kamiokande:2018ofw}.
We see much larger event rates than the neutrino factory cases thanks
to the large neutrino flux of the conventional beam.

\subsection{Background subtraction}
As we see above, we have background events from $\bar \nu_\mu \to \bar
\nu_\mu$ for the measurement of $N^{\nu_e \to \nu_\mu}$.
The method of the neutron tagging can in principle distinguish
$\nu_\mu$ from $\bar \nu_\mu$ but the detail efficiencies are not
known at this stage.
As a reference, the efficiency of the charge identification is about
70\% at SK-Gd, and it is going to be better at Hyper-Kamiokande. In
this study, we study the cases with various efficiencies of charge
identification ($C_{\rm id}$), such as the most optimistic choice $C_{\rm id}
= 100\%$, middle $C_{\rm id} = 70\%$, and the pessimistic one $C_{\rm id} =
0\%$. For simplicity, we assume the efficiencies to be independent of
the neutrino energy.
We further assume that, by the time of the T2HK experiment to be being
conducted, $C_{\rm id}$ is measured and known.
In this situation, by simply subtracting the known $\bar \nu_\mu$ flux
multiplied by the known misidentification rate, one can obtain the
pure event rate $N^{\nu_e \to \nu_\mu}$.

In this method, the oscillation probability, $P(\nu_e\to\nu_\mu)$, at
each energy bin can be defined as 
\begin{align}
    P (\nu_e\to\nu_\mu)={1 \over \kappa} \frac{\qty(\kappa N_{\rm far}^{\nu_e\to\nu_\mu}+(1-\kappa)N_{\rm far}^{\bar{\nu}_\mu\to\bar{\nu}_\mu})-(1-\kappa)N_{\rm far}^{\bar{\nu}_\mu\to\bar{\nu}_\mu} \big|_{\rm T2HK}}{\tilde{N}_{\rm near}^{\nu_e\to\nu_e}},
    \label{eq10}
\end{align}
and
\begin{align}
    \kappa &\equiv \frac{1+C_{\rm id}}{2}.
    %\ ,\ C_{id}:\text{efficiency of charge identification}.
    \label{eq11}
\end{align}
Here, $\tilde{N}_{\rm near}$ and $N_{\rm far}$ denote the number of events at the source and at the detector, respectively, while $\tilde{N}_{\rm near}$ is
corrected by the factors of the detector volume and distance so that $N_{\rm far} / \tilde{N}_{\rm near}$ gives the probabilities. $\tilde{N}_{\rm near}$ in Eq.~(\ref{eq10}) different from $N_{\rm near}$ in Eq.~(\ref{eq14}). The factor $\kappa$ represents
the probability to correctly identify $\nu_\mu$ for an actual
$\nu_\mu$ event, while $1-\kappa$ is the misidentification
probability.
Therefore, the combination, $\kappa N_{\rm
far}^{\nu_e\to\nu_\mu}+(1-\kappa)N_{\rm
far}^{\bar{\nu}_\mu\to\bar{\nu}_\mu}$, is the actual number of events
to be observed. On the other hand, $N_{\rm far}^{\bar \nu_\mu \to \bar
\nu_\mu}\big|_{\rm T2HK}$ is the expected number of background events
calculated based on the knowledge of $P(\bar \nu_\mu
\to \bar \nu_\mu)$ measured at the T2HK experiment.

For $C_{\rm id} = 0$, i.e., $\kappa = 0.5$, it is not sensible to perform
the charge identification analysis since it is just the same as the
random choices of the events, while we lose a half of the signal
events by the selection.
In the following analysis, when we take $C_{\rm id} = 0$, we just do not
perform the charge identification analysis and simply add the
background events, and subtract the estimated amount by using the T2HK
data, i.e., we take $\kappa = 1$ and $1-\kappa = 1$ in
Eq.~\eqref{eq10}.

\subsection{Statistical uncertainties}

Combining $P(\nu_e \to \nu_\mu)$ defined in Eq.~\eqref{eq10} with the
oscillation probability $P(\nu_\mu \to \nu_e)$ measured at the T2HK
experiment, one can now obtain the T-violation in Eq.~\eqref{eq7} for
each energy bin labeled by $j$ such as
\begin{align}
    P_j^{\rm TV} = P_j(\nu_e - \nu_\mu) 
%    - {(N^{\nu_\mu \to \nu_e}_{\rm far})_j \big|_{\rm T2HK}
%\over (N^{\nu_\mu \to \nu_\mu}_{\rm near})_j \big|_{\rm T2HK}},
    - P_j (\nu_\mu - \nu_e) \big|_{\rm T2HK},
    \label{eq13}
\end{align}
where the first term is defined in Eq.~\eqref{eq10} and the latter
term is the time reversal probability measured at the T2HK experiment.
We discuss here the $\chi^2$ analysis to estimate the expected
accuracy of the measurements.

In defining $\chi^2$, we take the CP phase $\delta$ as well as the
matter density $\rho$ as the input parameters, and discuss how well we
can measure $\delta$. Although the matter density of the earth,
$\rho$, may be inferred by some other methods, we simply take the
attitude that we do not know the actual value. By doing so, one can
demonstrate how $\rho$ affects the measurement of $\delta$ and how
T violation is insensitive to the matter profile.

For the rest of oscillation parameters, we take the following
reference values, which correspond to the arithmetic average of the
best fit points in Ref.~\cite{ParticleDataGroup:2022pth} from three
groups. In this study, we consider only the case of NO (Normal
Ordering) for illustration.
%table of reference values
\begin{table}[H]
    \centering
    \begin{tabular}{c|c|c|c|c}
    % \hline 
         $\mathit{\Delta} m_{21}^2/10^{-5}\ \mathrm{eV}$ 
         &$\mathit{\Delta} m_{31}^2/10^{-3}\ \mathrm{eV}$ 
         &$\theta_{12}$
         &$\theta_{13}$
         &$\theta_{23}$ \\
         \hline 
         $7.43$ & $2.432$ & $\ang{33.9}$ & $\ang{8.49}$ & $\ang{48.1}$ \\
        %  \hline 
    \end{tabular}
    \caption{The reference values of oscillation parameters for the normal mass ordering.}
    \label{tab1}
\end{table}

When we take the true values for the input parameters as $\delta_0$
and $\rho_0$, $\chi^2$ for the T-violation measurement as a function
of postulated values of $\delta = \delta^{\rm test}$ and $\rho =
\rho^{\rm test}$ is defined as
\begin{align}
    \chi^2_{\rm TV} (\delta^{\rm test}, \rho^{\rm test})
    &\equiv \sum_j \frac{\qty[ P^{\rm TV}_j\qty(\delta_0,\ \rho_0) - P^{\rm TV}_j\qty(\delta^{\rm test},\ \rho^{\rm test}) ]^2}{\qty(\mathit{\Delta}P^{T\rm V}_j)^2} \label{eq12} 
\end{align}
where $j$ runs over energy bins. Although we use the probability
difference $P^{\rm TV}_j$ as the observable in this definition, the
actual measurements are of course based on counting of $(\kappa N_{\rm
far}^{\nu_e \to \nu_\mu} + (1-\kappa) N_{\rm far}^{\bar \nu_\mu \to
\bar \nu_\mu})_j$ at the neutrino factory to Hyper-Kamiokande
experiment and $(N_{\rm far}^{\bar \nu_\mu \to \bar \nu_\mu})_j
\big|_{\rm T2HK}$ as well as $(N_{\rm far}^{\nu_\mu \to \nu_e})_j
\big|_{\rm T2HK}$ at the T2HK experiment as it is clear in the defining
equation in Eqs.~\eqref{eq10} and \eqref{eq13}.

Accordingly, the statistical error $\mathit{\Delta}P^{\rm TV}_j$ is
given by 
\begin{align}
    \qty(\mathit{\Delta} P^{\rm TV}_j)^2 &=\qty(\mathit{\Delta} P_j^{\nu_e\to\nu_\mu})^2 + \qty(\mathit{\Delta} P_j^{\nu_\mu\to\nu_e})^2 \notag \\
               &=  (P_j^{\nu_e\to\nu_\mu})^2\qty{ 
               {\qty(\frac{\mathit{\Delta} \qty(\kappa N^{\nu_e\to\nu_\mu}_{\rm far}+(1-\kappa)N^{\bar{\nu}_\mu\to\bar{\nu}_\mu}_{\rm far})}{\kappa N^{\nu_e\to\nu_\mu}_{\rm far}})_j^2} 
               + \qty(\frac{\mathit{\Delta}N^{\nu_e\to\nu_e}_{\rm near}}{N^{\nu_e\to\nu_e}_{\rm near}})_j^2 }  \notag \\
               &+ (P_j^{\nu_\mu\to\nu_e})^2\qty{ \qty(\frac{\mathit{\Delta}N^{\nu_\mu\to\nu_e}_{\rm far}}{N^{\nu_\mu\to\nu_e}_{\rm far}})^2_j + 	\qty(\frac{\mathit{\Delta}N^{\nu_\mu\to\nu_\mu}_{\rm near}}{N^{\nu_\mu\to\nu_\mu}_{\rm near}})_j^2 } \notag \\
               &= (P_j^{\nu_e\to\nu_\mu})^2\qty{ 
               {\qty(\frac{\sqrt{\qty(\kappa N^{\nu_e\to\nu_\mu}_{\rm far}+(1-\kappa)N^{\bar{\nu}_\mu\to\bar{\nu}_\mu}_{\rm far})}}{\kappa N^{\nu_e\to\nu_\mu}_{\rm far}})_j^2 }
               + \qty(\frac{\sqrt{N^{\nu_e\to\nu_e}_{\rm near}}}{N^{\nu_e\to\nu_e}_{\rm near}})_j^2 }  \notag \\
               &+ (P_j^{\nu_\mu\to\nu_e})^2\qty{ \qty(\frac{\sqrt{N^{\nu_\mu\to\nu_e}_{\rm far}}}{N^{\nu_\mu\to\nu_e}_{\rm far}})^2_j + 	\qty(\frac{\sqrt{N^{\nu_\mu\to\nu_\mu}_{\rm near}}}{N^{\nu_\mu\to\nu_\mu}_{\rm near}})_j^2 } \notag \\
               &\sim (P_j^{\nu_e\to\nu_\mu})^2
               {\qty(\frac{\sqrt{\qty(\kappa N^{\nu_e\to\nu_\mu}_{\rm far}+(1-\kappa)N^{\bar{\nu}_\mu\to\bar{\nu}_\mu}_{\rm far})}}{\kappa N^{\nu_e\to\nu_\mu}_{\rm far}})_j^2 }
               + (P_j^{\nu_\mu\to\nu_e})^2 \qty(\frac{\sqrt{N^{\nu_\mu\to\nu_e}_{\rm far}}}{N^{\nu_\mu\to\nu_e}_{\rm far}})^2_j
               \label{eq14}
\end{align}
Here, we assume that the number of events to obey the Gaussian
distribution, and the variances of each observables are simply added
to obtain the total uncertainty as each measurement is supposed to be
independent.
In Eq.~(\ref{eq14}), $N_{\rm near}$ is the value before applying
corrections for distance and detector volume. Note that $N_{\rm near}$ here is different from $\tilde N_{\rm near}$ in Eq.~\eqref{eq10} which is defined as the one 
after the corrections. This is because the
applied corrections are known, and the number of events observed is
before the corrections are applied. Therefore, the uncorrected number
of events should be used for error evaluation. In any case,
however, since $N_{\rm near}$ in Eq.~(\ref{eq14}) is much larger than
$N_{\rm far}$, the contribution of this part is negligible, and the
final expression does not contain $N_{\rm near}$.

\begin{figure}[H]
    \centering
    \includegraphics[width=15cm]{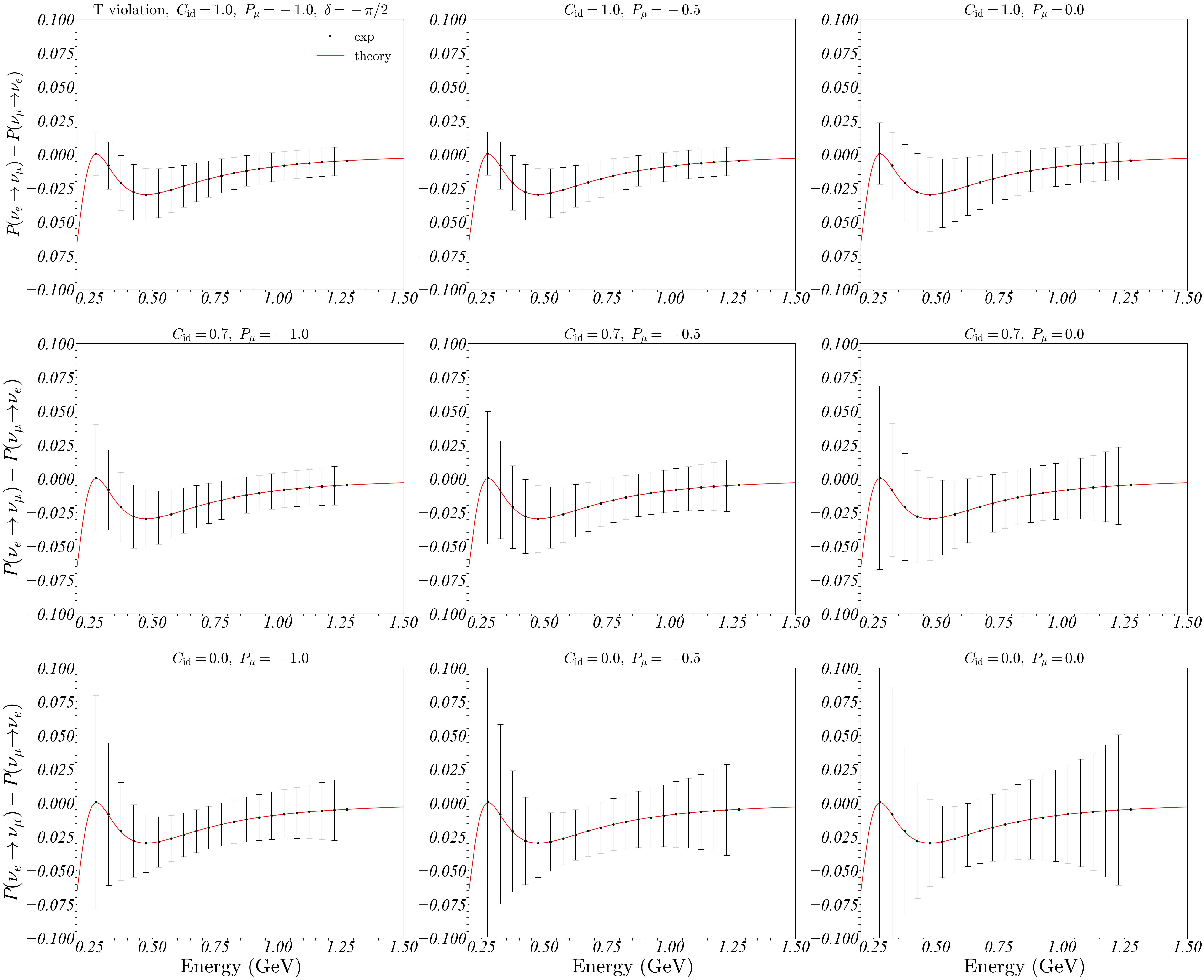}
    \caption{Reconstructed T violation from the expected neutrino flux
    with errors. The CP phase $\delta$ is set to be $-\pi/2$. From top to bottom,
    the graphs show the cases with the charge identification $C_{\rm id} = 1.0,\ 0.7$, and $0.0$, and
    from left to right, show the cases with the anti-muon polarization $P_\mu = -1.0,\ -0.5$, and
    $0.0$.}
    \label{fig5}
\end{figure}

\begin{figure}[H]
    \centering
    \includegraphics[width=15cm]{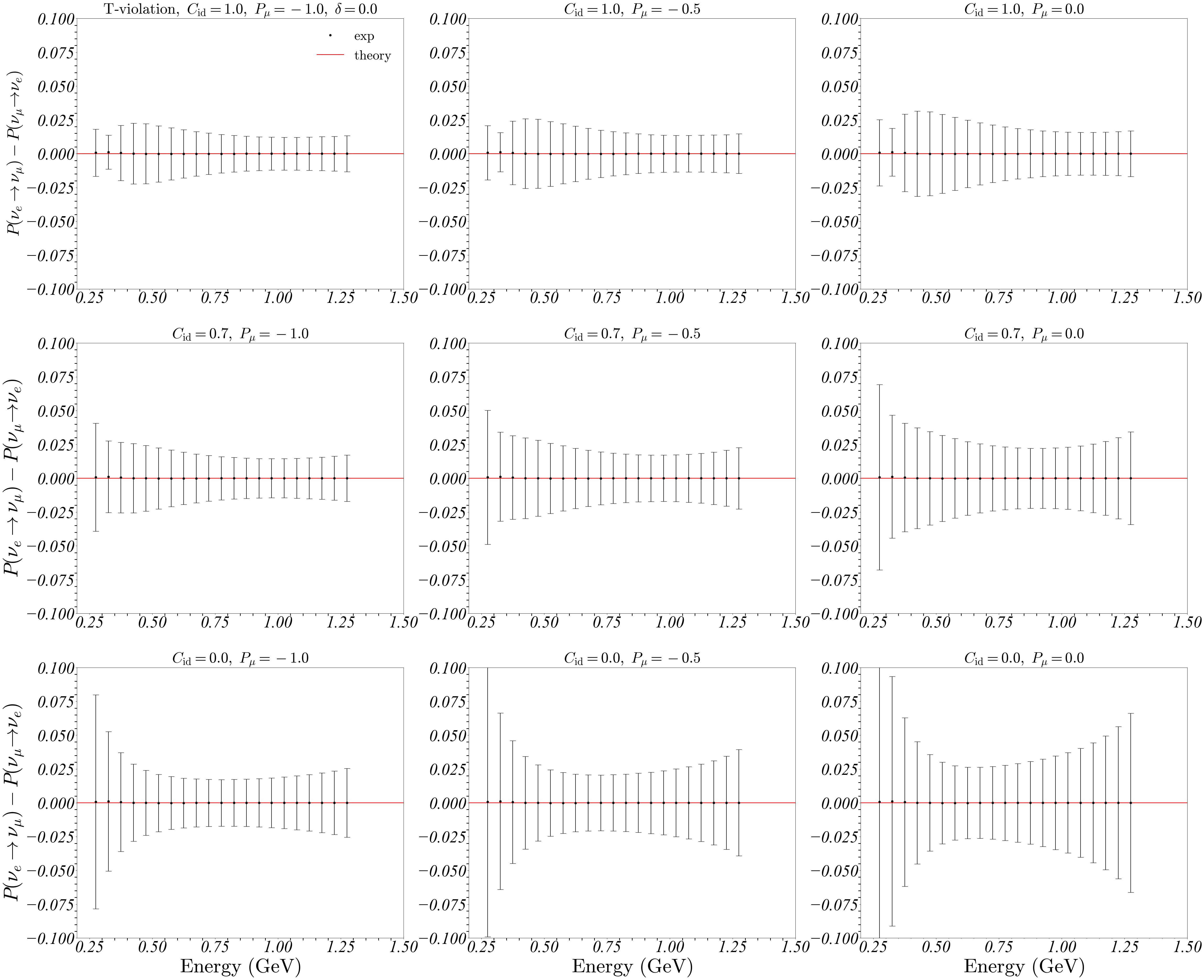}
    \caption{Reconstructed T violation from the expected neutrino flux with errors.The CP phase $\delta$ is set to be $0.0$.}
    \label{fig6}
\end{figure}

By using the above estimated errors in each energy bin, we present the
expected accuracy of the T-violation measurements in Fig.~\ref{fig5}
($\delta_0 = -\pi/2$, $\rho_0$ = 2.6~g/cm$^3$) and in Fig.~\ref{fig6} ($\delta_0
= 0.0$, $\rho_0$ = 2.6~g/cm$^3$). We take the bin size to be 50~MeV, and take the polarization factor $P_\mu$ to be $-1.0$, $-0.5$, and $0.0$ (from left to right).
The efficiency of the charge identification, $C_{\rm id}$ is taken to be
$1.0$, $0.7$ and $0.0$ (from top to bottom).


One can see that the T violation is maximized near $E_\nu \sim
500$~MeV where the background is suppressed by the neutrino
oscillation. This feature helps to give good sensitivity to
T violation. Also, even for zero efficiency of the charge
identification, $C_{\rm id} = 0.0$, the sensitivity is not much worse than
the case of the perfect identification $C_{\rm id} = 1.0$ as we will see
more explicitly later.

%%%%%%%%%%%%%%%%%%%%%%%%%%%%%%%%%%%%%%%%%%%%%%%%%%%%%%%%%%%%%%%%%
%\newpage
%\section{T-violation measurements}
%
\subsection{Parameter determination}
\begin{figure}[H]
    \centering
    \includegraphics[width=15cm]{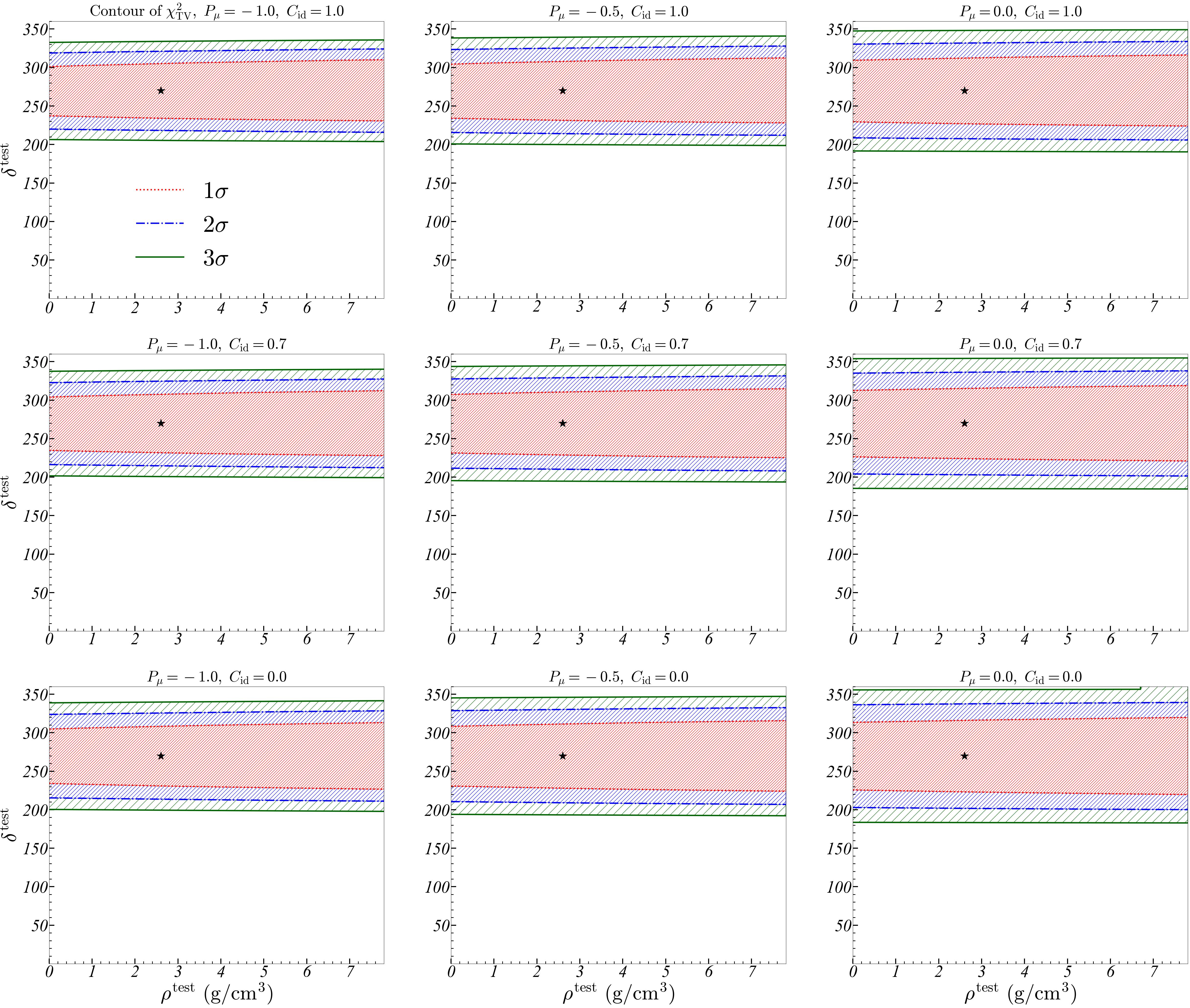}
    \caption{The contour of $\chi^2_{\rm TV}$ is defined in Eq.~(\ref{eq12}). The shaded regions represent $1\sigma$, $2\sigma$, and $3\sigma$ confidence interval, respectively, from inside to outside. From top to bottom, the figures show the cases with $C_{\rm id}=1.0,\ 0.7,\ 0.0$. From left to right, the figures show the cases with the anti-muon polarization $P_\mu = -1.0,\ -0.5,\ 0.0$. The stars in these figures represent the true input value ($\delta_0=-\pi/2,\ \rho_0=2.6\ \mathrm{g/cm^3}$).}
    \label{fig11}
\end{figure}

In Fig.~\ref{fig11}, we show the contour plots of $\chi^2_{\rm TV}
(\delta^{\rm test}, \rho^{\rm test})$. The shaded regions correspond
to $1\sigma$, $2\sigma$, and $3\sigma$ allowed regions. We set true
values as $\delta_0=-\pi/2$ and $\rho_0=2.6\ \mathrm{g/cm^3}$. 
Again, we take $P_\mu = -1.0$, $-0.5$, and $0.0$ from left
to right, and $C_{\rm id} = 1.0$, $0.7$, and $0.0$ from top to bottom.
The results are almost unchanged for different choices of $C_{\rm id}$.
This is quite important since the T-violation measurement can be
performed without having a good identification strategy between
$\nu_\mu$ and $\bar \nu_\mu$.
Independent of $C_{\rm id}$, CP (or T) conserving point, $\delta = 0$ and
$\pi$ can be excluded at the level of 3$\sigma$ even for the
unpolarized muon beam.
For the best case, $C_{\rm id} = 1.0$ and $P_\mu = -1.0$, the
CP angle $\delta$ is determined with an accuracy of about $\pm
\ang{30}$.

\begin{figure}[H]
    \centering
    \includegraphics[width=15cm]{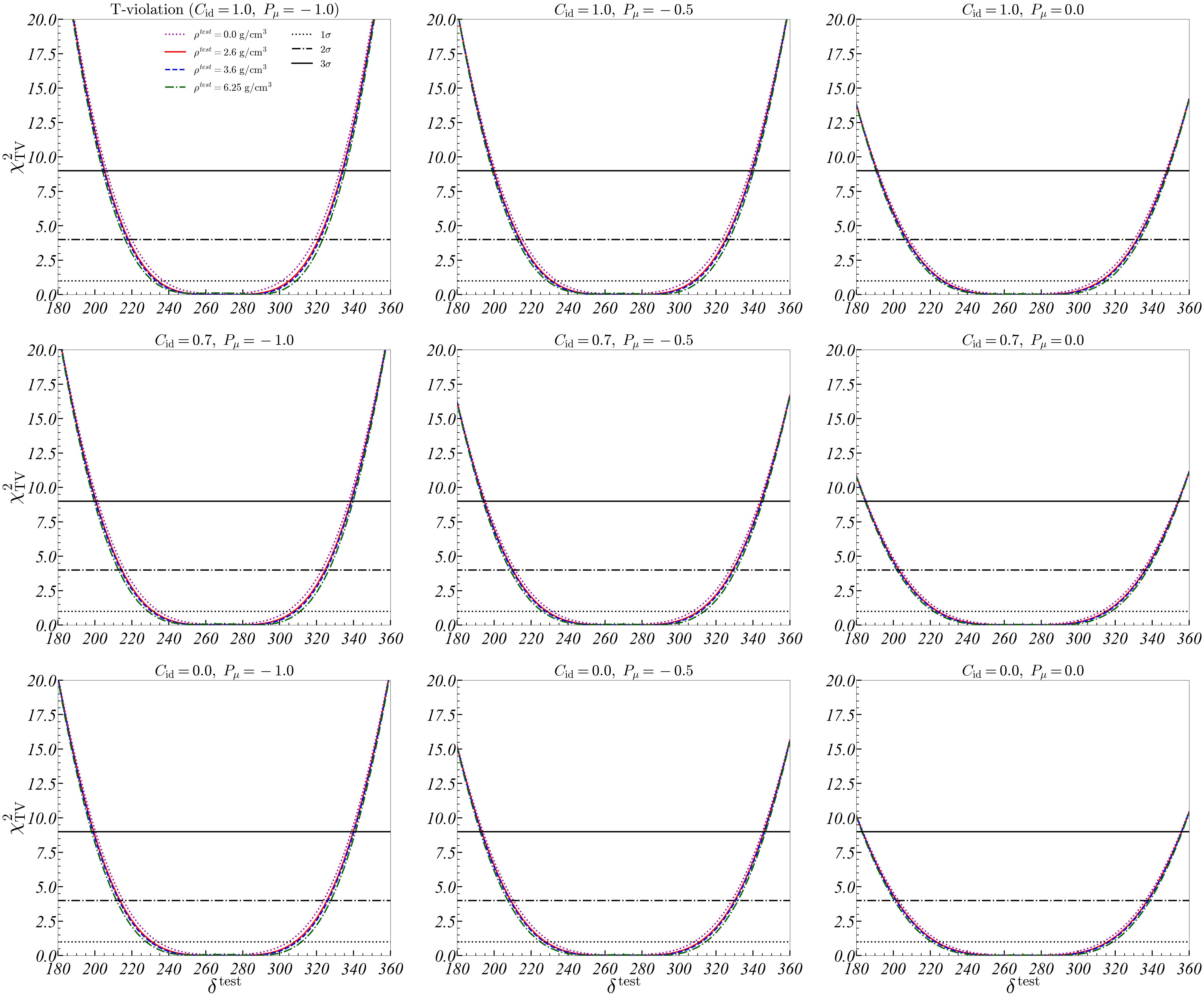}
    \caption{The values of $\chi^2_{\rm TV}$ in the region near $\delta=-\pi/2$. The purple dotted, red solid, blue dashed, and green dashdot lines represent $0.0~\mathrm{g/cm^3}$, $2.6~\mathrm{g/cm^3}$, $3.6~\mathrm{g/cm^3}$, and $6.25~\mathrm{g/cm^3}$, respectively. From top to bottom, the figures show the cases with $C_{\rm id}=1.0,\ 0.7,\ 0.0$. From left to right, the figures show the cases with anti-muon polarization $P_\mu=-1.0,\ -0.5,\ 0.0$. Three horizontal lines ($1\sigma$ dot, $2\sigma$ dashdot, $3\sigma$ solid) are values of $\chi^2$ for 1 d.o.f. The true input  values are set to be $\delta_0=-\pi/2$ and $\rho_0=2.6\ \mathrm{g/cm^3}$.}
    \label{fig13}
\end{figure}

It is clear from the figure that $\chi^2_{\rm TV}$ depend on
$\rho^{\rm test}$ only very slightly. 
This behavior is the obvious consequence of the time reversal symmetry
of the earth.
As we know that $\chi^2_{\rm TV}$ only depends on one of the
parameters, $\delta$, we define the confidence level as that for a
single degree of freedom, i.e., $\chi^2 = 1$ as 1$\sigma$.
For completeness, we show in Fig.~\ref{fig13} the values of
$\chi^2_{\rm TV}$ near the minima for choices of $\rho^{\rm test}$ as
$0.0$~g/cm$^3$, $2.6$~g/cm$^3$, $3.6$~g/cm$^3$, and $6.25$~g/cm$^3$
while fixing the true value to be $\rho_0 = 2.6$~g/cm$^3$.
% The zoomed
%up figure near the $\chi^2$ minimum is shown in Fig.~\ref{fig13}.

\begin{figure}[H]
    \centering
    \includegraphics[width=16cm]{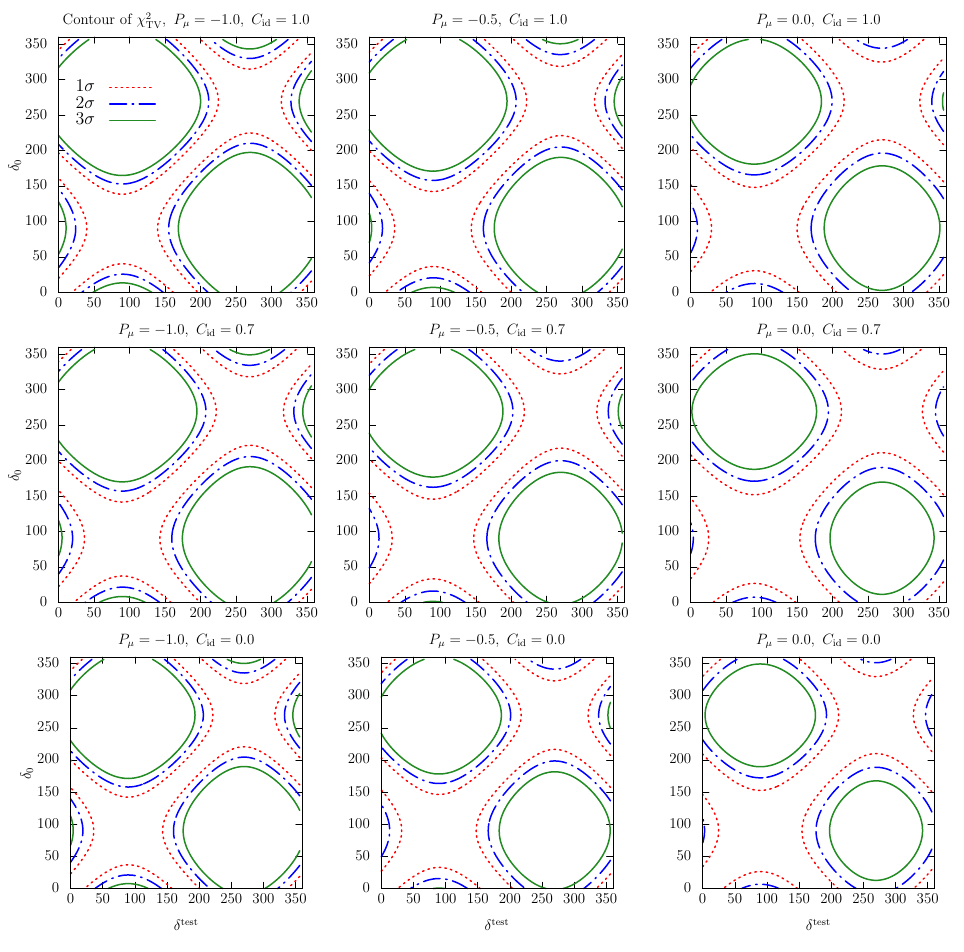}
    \caption{The confidence intervals on the $\delta^{\rm test}$ - $\delta_0$ plane for various choices of the muon polarization and the charge identification efficiencies. The matter density is fixed as $\rho_0 = \rho^{\rm test} = 2.6 $~g/cm$^3$. The red dot, blue dashed, and green solid lines correspond to $1\sigma$, $2\sigma$, and $3\sigma$ levels, respectively.}
    \label{fig-d-d0}
\end{figure}
% %fig of chi^2 contour

Finally, we show in Fig.~\ref{fig-d-d0} the confidence intervals on the $\delta^{\rm test}$ - $\delta_0$ plane for various choices of the muon polarization and the charge identification efficiencies.
The matter density is fixed as $\rho_0 = \rho^{\rm test} = 2.6 $~g/cm$^3$.
The cross sections of $\delta_0 = \ang{270}$ correspond to
Fig.~\ref{fig13}. One can read off the expected precision for each CP angle $\delta_0$. Conversely, one can also read off the required size of the CP angle for T violation to be measurable. For example, at the $3\sigma$ level, the T violation can be measured for $\ang{50} < \delta_0 < \ang{130}$ and $\ang{230} < \delta_0 < \ang{310}$, at the best case, $P_\mu = -1.0$ and $C_{\rm id} = 1.0$. 

%%%%%%%%%%%%%%%%%%%%%%%%%%%%%%%%%%%%%%%%%%%%%%%%%%%%%%%%%%%%%%%%%%%%%%
\subsection{$\chi^2$ analysis of CP violation at T2HK}

\begin{figure}[H]
    \centering
    \includegraphics[width=13cm]{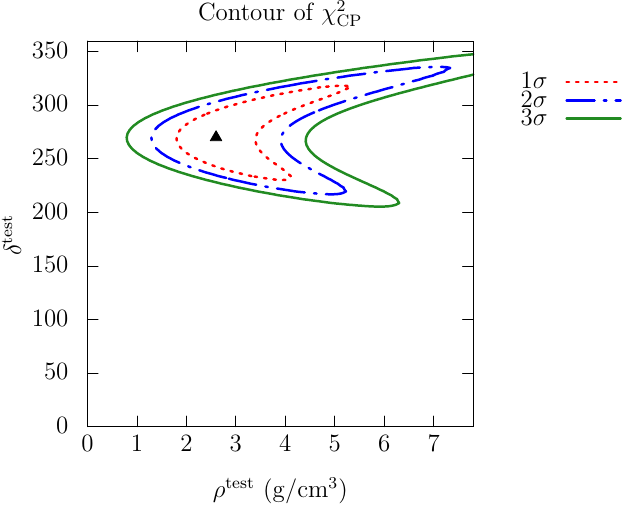}
    \caption{The contour of $\chi^2_{\rm CP}$. The red dot, blue
    dashdot, and green solid lines represent $1\sigma$, $2\sigma$, and
    $3\sigma$ confidence interval, respectively. The triangle symbol represent the true input value
    ($\delta_0=-\pi/2,\ \rho_0=2.6\ \mathrm{g/cm^3}$).}
    \label{fig10}
\end{figure}

In contrast to the T-violation measurement, CP violation suffers from
the uncertainties in the matter profile of the earth.
We consider CP violation,
\begin{align}
    P^{\rm CP}_j = P_j (\nu_\mu \to \nu_e) \big|_{\rm T2HK} - P_j (\bar \nu_\mu \to \bar \nu_e) \big|_{\rm T2HK}, 
\end{align}
and we define $\chi^2_{\rm CP}
(\delta^{\rm test}, \rho^{\rm test})$ in a similar way as T violation. 
We show the contour of the
confidence interval in Fig.~\ref{fig10}.

We see a non-trivial dependence of the allowed region on $\rho^{\rm
test}$, and it is clear that a good knowledge of the matter density
profile will be necessary for this measurement. The measurement of
T violation will be an important additional information for the
measurement of the CP angle $\delta$.

The precision to measure the CP angle $\delta$ at the T2HK obtained here is 
consistent with the value, $\Delta \delta \sim \ang{30}$, quoted in Ref.~\cite{Hyper-Kamiokande:2018ofw} under the condition that
we have perfect knowledge of the matter density.
As we see before, one can obtain a similar level of precision by the T violation analysis. However, the point of the T violation measurement would be the independent
measurement and the consistency check of the underlying assumptions such as 
the matter profile and the CPT theorem.

\subsection{Discussion: possible CPT test?}
Although we do not try in this paper, 
one would be able to perform a similar analysis for CPT violation,
\begin{align}
    P_j^{\rm CPT} = P_j (\nu_e \to \nu_\mu) - P_j (\bar \nu_\mu \to \bar \nu_e) \big|_{\rm T2HK} = P_j^{\rm TV} + P_j^{\rm CP}.
    \label{eq:cpt}
\end{align}
Our analyses up to the previous subsection are performed under the assumption of the standard three neutrino paradigm.
In this case, of course, CP violation and T violation are both measuring the same angle $\delta$, and thus $P_j^{\rm CPT} = 0$, trivially.
The points of measuring/observing T violation in future 
experiments are, however,   
an independent crosscheck of the ``previous'' results from T2HK with different systematics and more importantly a test of the CPT theorem by separately measuring the two terms in Eq.~\eqref{eq:cpt}, rather than improvement of the statistical precision to measure $\delta$.

Indeed, the neutrino mass is the smallest mass scale of the Standard Model, and its nature is totally unknown. Among many scenarios, there have been proposals 
of explaining the neutrino masses via small violation of the Lorentz invariance, for example, in Refs.~\cite{Coleman:1998ti, Cohen:2006ky}.
The analysis of this kind
will be a quite important fundamental test of symmetry in 
physical laws of the Universe.

%%%%%%%%%%%%%%%%%%%%%%%%%%%%%%%%%%%%%%%%%%%%%%%%%%%%%%%%%%%%%%%%%%%%
\section{Summary}
\label{sec:summary}

Motivated by the recent discussions of the intense muon beam for muon
colliders, we studied the possibility of measuring T violation,
$P(\nu_e \to \nu_\mu) - P(\nu_\mu - \nu_e)$ by assuming the baseline
from J-PARC to Hyper-Kamiokande. We take a scenario that a $\mu^+$
beam will be available at J-PARC as proposed in
Ref.~\cite{Hamada:2022mua}, so that a large flux of $\nu_e$ will be
pointing towards Hyper-Kamiokande.
By combining with the measurement at the T2HK experiment which uses
the conventional $\nu_\mu$ beam, the T violation can be defined.
The T violation, the probability difference above, is a quantity that
does not suffer from the uncertainty of matter density of the earth.

Under reasonable assumptions on the muon intensity, the results show
that the observation of T violation is possible if $|\sin \delta|$ is
large enough. 
For the case of maximum CP violation, i.e., $\delta = -\pi/2$, one can
exclude the CP invariant theory $\delta = 0$ or $\pi$ at more than
$3\sigma$ level.
There are two important parameters in the experimental set-up, which
are the beam polarization of muons and the efficiency of the charge
identification to distinguish $\nu_\mu$ from $\bar \nu_\mu$ at
Hyper-Kamiokande.
There is a fortunate fact that the background $\bar \nu_\mu$ events
from the muon decays are suppressed at the most important energy range
for T violation by setting the muon beam energy to give the
oscillation maximum. Even with no ability of distinguish $\nu_\mu$ and
$\bar \nu_\mu$ events, the sensitivities do not change significantly.
Rather, the beam polarization to increase the forward $\nu_e$ will be
more important at the actual experiment.

We considered $\chi^2$ for CP violation and compared it with that for
T violation. For CP violation, $\chi^2_{\rm CP}$ depends significantly
on both $\delta$ and $\rho$, whereas for T violation, $\chi^2_{\rm
TV}$ is almost independent on $\rho$. This means T violation would not
suffer from the uncertainties in the matter density profile of the
earth, and give a clearer signal.

Testing T violation in the lepton sector has not been conducted yet,
so it remains as an important task in particle physics. In this study, we demonstrated
the feasibility for the measurement, while we considered only the
statistical error. A more complete analysis will be necessary to
establish the feasibility.
Once it is measured, it is going to be a very nontrivial test of the
three neutrino paradigm or possibly of the CPT theorem which is one of
the most fundamental features in the quantum field theory.

% In addition, the new channel of
%T-violation can contribute to the global analyses of $\delta$. 

\section*{Acknowledgements}
We would like to thank Osamu Yasuda for useful discussions and 
Ken Sakashita and Takasumi Maruyama for useful
information on neutrino detectors. This work was in part supported by
JSPS KAKENHI Grant Numbers JP22K21350 (R.K.), JP21H01086 (R.K.),
JP19H00689 (R.K.), and JST SPRING Japan Grant Number JPMJSP2178.

%%%%%%%%%%%%%%%%%%%%%%%%%%%%%%%%%%%%%%%%%%%%%%%%%%%%%%%%%%%%%%%%%%%%

\appendix

\bibliographystyle{./utphys.bst}
\bibliography{./T-violation_ver2.bib}

\end{document}